\documentclass[pra,showpacs,showkeys,twocolumn]{revtex4}
\usepackage{graphicx,amsmath,amssymb}
\usepackage{epsf,color,graphicx}
\usepackage{pslatex,floatflt,latexsym}
\newcommand{\rmi}{\text{i}}

\newcommand{\dleft}{\left.\left}
\newcommand{\dright}{\right.\right}
\newcommand{\drangle}{\right>\!\!\right>}
\newcommand{\dlangle}{\left<\!\!\left<}
\begin{document}
\title{Generation of Entangled $N$-Photon States in a Two-Mode Jaynes--Cummings Model}
\author{C.~Wildfeuer} 
\author{D.~H.~Schiller} 
\affiliation{Fachbereich Physik, Universit\"{a}t Siegen, 
 D-57068 Siegen, Germany}
\email{wildfeuer@physik.uni-siegen.de}
\begin{abstract}
We describe a mathematical solution for the generation of entangled $N$-photon
states in two field modes. A simple and compact solution is presented for
a two-mode Jaynes--Cummings model by combining the two field modes in a way that only one of the two resulting
quasi-modes enters in the interaction term.
The formalism developed is then applied to calculate various generation probabilities analytically. We show that
entanglement, starting from an initial field and an atom in one defined state
may be obtained in a single step. We also show that entanglement may be built
up in the case of an empty cavity and excited atoms whose final states are
detected, as well as in the case when the final states of the initially excited atoms are not detected.
\end{abstract}
\maketitle
\section{Introduction}
Entangled states are one of the building blocks in quantum information
processing and non-locality tests \cite{Bouwmeester}. 
They can be used, in the case of the electromagnetic field, to improve
the sensitivity of interferometric measurements
\cite{Yurke,Hillery,Brif,Dowling}, and may help to overcome the classical
Rayleigh diffraction limit in quantum optical lithography \cite{Boto}.
A feasible way to generate such states is given by the atom-field interaction
in the framework of one- or two-mode Jaynes--Cummings (JC) models \cite{Deb,IkramZubairy,Rauschenbeutel,Solano,Fiurasek,Unanyan}. 
\par Here we consider the generation of entangled two-mode field states by different schemes inspired partly by Refs.~\cite{Schleich,Deb,IkramZubairy}.
We let two-level atoms interact, one at a time, with two degenerate
modes of a lossless cavity. Solving the corresponding JC model
algebraically by an SU(2) transformation, we discuss the generation of
entangled $N$-photon states of the general form
\begin{equation}\label{quasiBell}
        |\Psi_N\rangle=\sum_{k=0}^N c^{(N)}_k|N-k,k\rangle,
        \end{equation}  
which comprises the maximally entangled Bell states
 \begin{equation}\label{Bell}
   |\Psi_N^\pm\rangle = 
        \frac{1}{\sqrt{2}}\left(|N,0\rangle \pm |0,N\rangle\right).
   \end{equation}
The field states are defined in terms of the usual two-mode Fock states $|n_1,n_2\rangle:=|n_1\rangle_1|n_2\rangle_2$, with
        $n_1$ ($n_2$) photons in mode one (two). The two modes have the same
        energy and are in resonance with the two-level atom.
We solve the model algebraically by combining the two field modes into
two quasi-modes of which only one enters in the interaction term, yielding an
effective  one-mode JC model \cite{Quattropani}. Using its known solution and the transformation between mode and quasi-mode Fock
states, the generation probabilities of the entangled states are found for
        three different schemes. 
\section{Algebraic Solution of the Two-Mode Jaynes-Cummings Model}
\par The JC Hamiltonian for resonant interaction of a two-level
atom ($|e\rangle,|g\rangle$) with two field modes ($a_1, a_2$) in the dipole
and rotating wave approximation is given by
$H=H_0+H_{\text{int}}$, where 
\begin{eqnarray}
        H_0&=&\hbar\omega\left(\frac{\sigma_z+\mathbf{1}}{2}+\left(a_1^\dagger a_1 +
        a_2^\dagger a_2\right)\mathbf{1}\right),\\
        H_{\text{int}}&=&\hbar\left(\sigma^+\left(g_1 a_1 + g_2 a_2\right) 
        + \sigma^-\left(g_1^\ast a_1^\dagger + g_2^\ast
        a_2^\dagger\right)\right).
\end{eqnarray}
Here $\sigma_z:=|e\rangle\langle e|-|g\rangle\langle g|$,
  $\sigma^+:=|e\rangle\langle g|$, $\sigma^{-}:=|g\rangle\langle e|$ and
  $\mathbf{1}=|e\rangle\langle e|+|g\rangle\langle g|$ are operators for the two-level
  atom, $g_i$ is the coupling constant
  of the $i$th mode with the atom, and $\hbar\omega$ is the photon energy.
We introduce the {\em quasi-mode} operators
\begin{equation}
A_1=\gamma_1a_1+\gamma_2a_2,\quad A_2=-\gamma_2^\ast a_1 + \gamma_1^\ast a_2,\label{SU2}
\end{equation}
where $\gamma_i:=g_i/g$, and $g:=\sqrt{|g_1|^2+|g_2|^2}$.
Equation~(\ref{SU2}) defines an SU(2) transformation of the mode operators
$a_1,a_2$, leaving the commutation relations and the number-sum operator $a_1^\dagger
a_1+a_2^\dagger a_2=A_1^\dagger A_1 + A_2^\dagger A_2$ invariant.
The transformed Hamiltonian then reads
\begin{eqnarray}
  H_0&=&\hbar\omega\left(\frac{\sigma_z+\mathbf{1}}{2}+\left(A_1^\dagger A_1 +
  A_2^\dagger A_2\right)\mathbf{1}\right),\\
  H_{\text{int}}&=&\hbar g \left(\sigma^+A_1 + \sigma^-A_1^\dagger\right),
\end{eqnarray}
representing a JC Hamiltonian for the quasi-mode $A_1$ decoupled from the non-interacting
quasi-mode $A_2$.
Since $H_{\text{int}}$ depends only on quasi-mode one and $[H_0,H_{\text{int}}]=0$, the time
evolution operator $U(t)=\exp(-\rmi H_{\text{int}}t/\hbar)$ in the interaction picture is the
same as for a one-mode JC model. Expanding $U$ in the atom basis $\{|e\rangle,|g\rangle\}$ 
\begin{equation}\label{U1}
   U= U_{ee}|e\rangle\langle e|+U_{ge}|g\rangle\langle e| + U_{eg}|e\rangle\langle g| + U_{gg}|g\rangle\langle g|,
\end{equation}
the matrix elements $U_{ab}(t)$ are given by \cite{Zubairy}
\begin{eqnarray}\label{U2}
 U_{ee}&=&\cos\left(\tau\sqrt{A_1^\dagger A_1 +1}\right),\,\,\, U_{ge}=A_1^\dagger\frac{\sin\left(\tau\sqrt{A_1^\dagger A_1 +1}\right)}{\rmi\sqrt{A_1^\dagger A_1
  +1}},\nonumber\\U_{eg}&=&\frac{\sin\left(\tau\sqrt{A_1^\dagger A_1 +1}\right)}{\rmi\sqrt{A_1^\dagger A_1
  +1}}\,A_1,\,\,\, U_{gg}=\cos\left(\tau\sqrt{A_1^\dagger A_1}\right),\nonumber\\
\end{eqnarray}
where $\tau:=gt$ is the dimensionless ``interaction time''.
The model can be solved in the usual way in terms of quasi-mode Fock
states defined as the common eigenstates of $A_1^\dagger A_1$ and $A_2^\dagger A_2$. The complete solution is then found by giving the
relation between the quasi-mode and the mode Fock states.
\par The quasi-mode operators $A_i$, $A_i^\dagger$,
$i=1,2$, obey the same algebra as the mode operators $a_i$, $a_i^\dagger$, so
that two-quasi-mode Fock states (denoted by a double-ket) can be defined by
\begin{equation}\label{quasifock}
\dleft|n_1,n_2\drangle
:=\frac{{A_1^\dagger}^{n_1}{A_2^\dagger}^{n_2}}{\sqrt{n_1!n_2!}}\dleft|0,0\drangle.
\end{equation}
To find the transformation between the two-mode Fock states $|n_1,n_2\rangle$ and
the two-quasi-mode Fock states $\dleft|n_1,n_2\drangle$, we use Schwinger's
oscillator model \cite{Sakurai} and introduce angular momentum states
$|j,m\rangle$ and $\dleft|j,m\drangle$, where $j=(n_1+n_2)/2$ and
$m=(n_1-n_2)/2$. In cases where it is not obvious, we shall write a subindex S
on the state vectors to indicate the Schwinger angular momentum basis, e.g., $|2,0\rangle=|1,1\rangle_\text{S}$. 
Inserting Eq.~(\ref{SU2}) into Eq.~(\ref{quasifock}) and identifying the two
vacua $\dleft|0,0\drangle$ and $|0,0\rangle$, we obtain
\[\dleft|j,m\drangle
:=\frac{{\left(\gamma_1^\ast a_1^\dagger+\gamma_2^\ast a_2^\dagger\right)}^{j+m}{\left(-\gamma_2 a_1^\dagger + \gamma_1 a_2^\dagger\right)}^{j-m}}{\sqrt{(j+m)!(j-m)!}}\left|0,0\right\rangle.\]
Expanding the products, rearranging the terms \cite{Wigner} and using the
definition of the Fock basis $|n_1,n_2\rangle$ in terms of $a_1^\dagger$ and $a_2^\dagger$, we obtain the
important relation between the quasi-mode and the mode Fock bases  
\begin{eqnarray}
  \dleft|j,m\drangle&=&\sum_{m'=-j}^j D_{m'm}^{(j)}(\varphi,\vartheta,\chi)\left|j,m'\right\rangle,\label{quasito}\\ 
  |j,m\rangle&=&\sum_{m'=-j}^j D_{m'm}^{(j)^\dagger}(\varphi,\vartheta,\chi)\dleft|j,m'\drangle.\label{quasiback}
\end{eqnarray} 
Here $D_{m'm}^{(j)}(\varphi,\vartheta,\chi)=\mathrm{exp}[-\rmi(m'\varphi +
  m\chi)]d_{m'm}^{(j)}(\vartheta)$ are the Wigner $D$-matrix elements of the SU(2)
  group \cite{Sakurai,Wigner}, with arguments determined by $\varphi=\varphi_1-\varphi_2$, $\chi=\varphi_1+\varphi_2$, $\cos(\vartheta/2):=|\gamma_1|$, 
 $\sin(\vartheta/2):=|\gamma_2|$, and $\gamma_i=|\gamma_i|\mathrm{exp}(\rmi\varphi_i)$.
It follows that the mode and
  quasi-mode Fock states belonging to the same total number of photons,
  $n_1+n_2=2j$, are related by an irreducible rotation matrix of weight $j$
  and with Euler angles determined solely by the
  interaction constants. 
\par The action of $U_{ab}$ on the field states is easily calculated in the
  quasi-mode Fock basis
\begin{eqnarray}\label{quasiU}
   U_{ee}(\tau)\dleft|j,m\drangle&=&\cos\left(\tau\sqrt{j+m+1}\right)\dleft|j,m\drangle,\nonumber\\  
   U_{ge}(\tau)\dleft|j,m\drangle&=&-\rmi\sin\left(\tau\sqrt{j+m+1}\right)\dleft|j+{\textstyle\frac{1}{2},m+\frac{1}{2}}\drangle,\nonumber\\
   U_{eg}(\tau)\dleft|j,m\drangle&=&-\rmi\sin\left(\tau\sqrt{j+m}\right)\dleft|{\textstyle
   j-\frac{1}{2},m-\frac{1}{2}}\drangle,\nonumber\\
   U_{gg}(\tau)\dleft|j,m\drangle&=&\cos\left(\tau\sqrt{j+m}\right)\dleft|j,m\drangle,
 \end{eqnarray}
showing that $U_{ee}$ and $U_{gg}$ do not change the number of quasi-photons,
whereas $U_{ge}$ ($U_{eg}$) act as creation (annihilation) operators of
quasi-mode one.
Using Eq.~(\ref{quasito}) and Eq.~(\ref{quasiback}), we find for the action on
the usual Fock states
 \begin{eqnarray}
   U_{ee}(\tau)\left|j,m\right\rangle&=&\sum_{m'=-j}^j C_{m'm}^j(\tau)\left|j,m'\right\rangle,\nonumber\\  
   U_{ge}(\tau)\left|j,m\right\rangle&=&\sum_{m'=-j-\frac{1}{2}}^{j+\frac{1}{2}}S_{m'm}^j(\tau)\left|j+{\textstyle\frac{1}{2}},m'\right\rangle,\nonumber\\
   U_{eg}(\tau)\left|j,m\right\rangle&=&\sum_{m'=-j+\frac{1}{2}}^{j-\frac{1}{2}}\overline{S}_{m'm}^j(\tau)\left|j-{\textstyle\frac{1}{2}},m'\right\rangle,\nonumber\\
   U_{gg}(\tau)\left|j,m\right\rangle&=&\sum_{m'=-j}^j\overline{C}_{m'm}^j(\tau)\left|j,m'\right\rangle,
 \end{eqnarray}
where we have introduced the following coefficients
 \begin{eqnarray}
   C_{m'm}^j(\tau)&=&\sum_{\nu=-j}^j\cos\left(\tau\sqrt{j+\nu+1}\right)D_{m'\nu}^{(j)}D_{\nu m}^{(j)^\dagger},\nonumber\\
  S_{m'm}^j(\tau)&=&-\rmi\sum_{\nu=-j}^j\sin\left(\tau\sqrt{j+\nu+1}\right)D_{m',\nu+\frac{1}{2}}^{(j+\frac{1}{2})}D_{\nu m}^{(j)^\dagger},\nonumber\\
  \overline{S}_{m'm}^j(\tau)&=&-\rmi\sum_{\nu=-j}^j\sin\left(\tau\sqrt{j+\nu}\right)D_{m',\nu-\frac{1}{2}}^{(j-\frac{1}{2})}D_{\nu m}^{(j)^\dagger},\nonumber\\
 \overline{C}_{m'm}^j(\tau)&=&\sum_{\nu=-j}^j\cos\left(\tau\sqrt{j+\nu}\right)D_{m'\nu}^{(j)}D_{\nu m}^{(j)^\dagger}.
\end{eqnarray}
Given the above equations, we now have all the ingredients to calculate the time
evolution of the density operator according to $\rho(t)=U(t)\rho(0)U^\dagger(t)$.
\section{Generation of Entanglement in one Step}
 \par We start with the calculation of the probability to find at time $\tau$ the field state
 $|\Psi_N\rangle$ in Eq.~(\ref{quasiBell}), assuming an initial field state $|\xi\rangle$ and an atom
 entering the cavity in either the excited or ground state.
The analytical calculation is straightforward.
The initial field state is expanded according to
\begin{equation}\label{initial}
\left|\xi\right\rangle=\sum_{n_1=0}^\infty\sum_{n_2=0}^\infty
b_{n_1n_2}\left|n_1,n_2\right\rangle
=\mathop{\sum\nolimits'}\limits_{j=0}^\infty\sum_{m=-j}^j\tilde{b}_{jm}\left|j,m\right\rangle,
\end{equation}
where the primed summation symbol indicates a sum over integer
and half integer values of $j$. The expansion coefficients with respect to the
Fock and Schwinger basis are related by $b_{j+m,j-m}=\tilde{b}_{j,m}$.
The state $|\Psi_N\rangle$ is given in the Schwinger
basis by 
\begin{equation}
\left|\Psi_N\right\rangle=\sum_{m=-\frac{N}{2}}^{N/2}\tilde{c}_{\frac{N}{2}\,m}\left|{\textstyle\frac{N}{2}},m\right\rangle_\text{S},
\end{equation} 
with density operator $\rho_{\Psi_N}={|\Psi_N\rangle\langle \Psi_N|}$.
From the time-evolved initial states
\begin{eqnarray*}U\left|e;\xi\right\rangle&=&U_{ee}\left|e;\xi\right\rangle+U_{ge}\left|g;\xi\right\rangle,\\ 
U\left|g;\xi\right\rangle&=&U_{eg}\left|e;\xi\right\rangle+U_{gg}\left|g;\xi\right\rangle,
\end{eqnarray*} 
we obtain the reduced density operator of the field by tracing out the atomic degrees of freedom:
$\rho^{(a)}_{\text{F}}(t)={\text{tr}}_{\text{A}}\left(U(t)|a;\xi\rangle\langle
a;\xi|U^\dagger(t)\right)$, $a=e$ or $g$. 
The probability to find $|\Psi_N\rangle$ at time $t$ follows from
$\langle \rho_{\Psi_N}^{(a)}\rangle =
{\text{tr}}(\rho_{\text{F}}^{(a)}(t)\rho_{\Psi_N})$ and is
given by 
\begin{eqnarray}
     \left\langle\rho_{\Psi_N}^{(e)}\right\rangle
   &=&\left|\sum_{m=-\frac{N}{2}}^{N/2}\sum_{m'=-\frac{N}{2}}^{N/2}
     \tilde{b}_{\frac{N}{2}m}{\tilde c}^\ast_{\frac{N}{2}m'}C_{m'
     m}^{\frac{N}{2}}(\tau)\right|^2\nonumber\\
      &&{}+\left|\sum_{m=-\frac{N-1}{2}}^{(N-1)/2}\sum_{m'=-\frac{N}{2}}^{N/2}
     \tilde{b}_{\frac{N-1}{2}m} {\tilde c}_{\frac{N}{2}m'}^\ast S_{m'
     m}^{\frac{N-1}{2}}(\tau)\right|^2\label{singlestepex}
  \end{eqnarray}
for the initial atom-field state $|e;\xi\rangle$, and by
\begin{eqnarray}
    \left\langle\rho_{\Psi_N}^{(g)}\right\rangle
   &=&\left|\sum_{m=-\frac{N}{2}}^{N/2}\sum_{m'=-\frac{N}{2}}^{N/2}
     \tilde{b}_{\frac{N}{2}m}{\tilde c}_{\frac{N}{2}m'}^\ast{\overline
     C}^\frac{N}{2}_{m' m}(\tau)\right|^2\nonumber\\
    &&{}+\left|\sum_{m=-\frac{N+1}{2}}^{(N+1)/2}\sum_{m'=-\frac{N}{2}}^{N/2}
     \tilde{b}_{\frac{N+1}{2}m}
     {\tilde c}_{\frac{N}{2}m'}^\ast {\overline
     S}_{m' m}^\frac{N+1}{2}(\tau)\right|^2\label{singlestepground}
\end{eqnarray}
for the initial state $|g;\xi\rangle$.
It follows that in order to obtain non-vanishing probabilities at time
$\tau$, the initial field state must contain at least one of the Fock states
from the set
 \begin{eqnarray}
   \lefteqn{\{|N,0\rangle,|N-1,1\rangle,\ldots,|0,N\rangle\}}\nonumber\\
  &&\cup\{|N-1,0\rangle,|N-2,1\rangle,\ldots,|0,N-1\rangle\},\label{set1}
 \end{eqnarray} 
if the atom is initially in the excited state, or from the set
\begin{eqnarray}
\lefteqn{\{|N,0\rangle,|N-1,1\rangle,\ldots,|0,N\rangle\}}\nonumber\\
&&\cup\{|N+1,0\rangle,|N,1\rangle,\ldots,|0,N+1\rangle\},\label{set2}
\end{eqnarray} 
if it is in the ground state. 
\par
\begin{figure}[htb]
\includegraphics{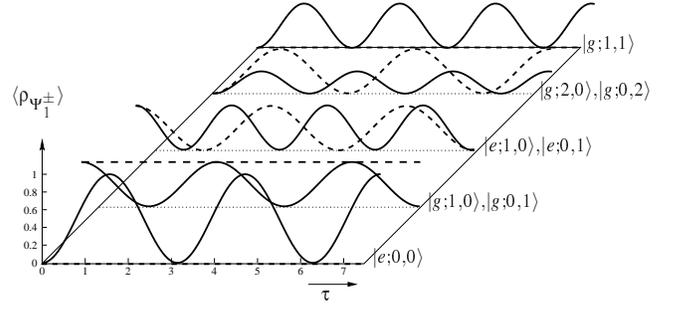}%
\caption{\label{figure1}Parametric plot of the generation probabilities
  $\langle\rho_{\Psi_1^+}\rangle$ (solid) and $\langle\rho_{\Psi_1^-}\rangle$
  (dashed) as function of time $\tau=g t$, for different initial atom-field
  states shown at the right.}
\end{figure}

 For $N=1$ the set of contributing atom-field states according to Eq.~(\ref{set1}) is 
 \begin{equation}
   \left\{|e;1,0\rangle,|e;0,1\rangle\right\}\cup\left\{|e;0,0\rangle\right\}\label{set1b}
 \end{equation}
and according to Eq.~(\ref{set2}) 
 \begin{equation}\left\{|g;1,0\rangle,|g;0,1\rangle\right\}\cup\left\{|g;2,0\rangle,|g;1,1\rangle,|g;0,2\rangle\right\}.\label{set2b}\end{equation}
We illustrate the probabilities for the generation of  
$|\Psi_1^\pm\rangle=(|1,0\rangle\pm|0,1\rangle)/\sqrt{2}$ in
Fig.~\ref{figure1}, where we have taken $g_1=g_2$, $\varphi_1=\varphi_2=0$
(real coupling constants).
The states shown on the right are just the initial atom-field states from
Eqs.~(\ref{set1b}) and (\ref{set2b}). The interesting case is $|e;0,0\rangle$, where the state $|\Psi_1^+\rangle$ is produced
periodically with probability one at the times $\tau_n=(n+1/2)/\pi$, for $n=0,1,\ldots$
\par Next we consider the creation of various Bell states, Eq.~(\ref{Bell}),
resulting from the initial atom-field state
$|e;N,0\rangle=\left|e;{\textstyle\frac{N}{2},\frac{N}{2}}\right\rangle_\text{S}$.
We obtain from Eq.~(\ref{singlestepex}) the entangled $N$-photon field states
$\left|\Psi_N^\pm\right\rangle$ with probabilities
\begin{equation}\label{ex}
\left\langle\rho^{(e)}_{\Psi_N^\pm}\right\rangle=\frac{1}{2}\left|C_{\frac{N}{2}\frac{N}{2}}^\frac{N}{2}(\tau)\pm
  C_{-\frac{N}{2}\frac{N}{2}}^\frac{N}{2}(\tau)\right|^2,
\end{equation}
as well as the entangled $(N+1)$-photon states $\left|\Psi_{N+1}^\pm\right\rangle$ with probabilities
\begin{equation}\label{gnd}
  \left\langle\rho^{(e)}_{\Psi_{N+1}^\pm}\right\rangle=\frac{1}{2}\left|S_{\frac{N+1}{2}\frac{N}{2}}^\frac{N}{2}(\tau)\pm
  S_{-\frac{N+1}{2}\frac{N}{2}}^\frac{N}{2}(\tau)\right|^2.
\end{equation}
In the case of Eqs.~(\ref{gnd}) the Bell states
$|\Psi_{N+1}^\pm\rangle$ have no overlap
with the initial field state $|N,0\rangle$.
The probabilities at time $\tau$, however, may come close to one for some particular
values of the coupling constants and interaction time. In this case we may
say that $|\Psi_{N+1}^\pm\rangle$ have been generated in a {\em single step} or
single shot. 
This property of the two-mode JC model can be understood if we think of the atom (re)emitting photons into and
(re)absorbing photons from the two modes many times during the interaction time
$\tau$. 
\section{Conditional Generation}
\par Next we present a {\em conditional} scheme for the generation of $N$-photon entangled states starting
with an empty cavity \cite{Deb,IkramZubairy}. The scheme implies sending
consecutively atoms in the excited state through a two-mode cavity and detecting
them in the ground state. We start with an initial atom-field
state $|e;0,0\rangle=\dleft|e;0,0\drangle_\text{S}$ and let the first atom
interact for a time $\tau_1$.
By using Eq.~(\ref{quasiU}) we obtain the state
\[
  U(\tau_1)\dleft|e;0,0\drangle_\text{S}=\cos{(\tau_1)}\dleft|e;0,0\drangle_\text{S}
  -\rmi \sin{(\tau_1)}\dleft|g;{\textstyle\frac{1}{2},\frac{1}{2}}\drangle_\text{S}.
\]
Detecting the atom in the ground state leaves the field in the state
$|\chi_1\rangle=K_1(-\rmi) \sin{(\tau_1)}\dleft|{\textstyle\frac{1}{2}},\frac{1}{2}\drangle_\text{S}$,
where $K_1=\left|\sin{(\tau_1)}\right|^{-1}\mathrm{exp}(\rmi\alpha_1)$ is a normalization constant.
By choosing the phase $\alpha_1$ appropriately, the factor entering the
normalized state may be set equal to one, yielding the state 
$\dleft|\frac{1}{2},\frac{1}{2}\drangle_\text{S}$. Proceeding this way the
field state obtained after $N$ conditional steps is simply given by  
\begin{equation}|\chi_N\rangle=\dleft|{\textstyle\frac{N}{2},\frac{N}{2}}\drangle_\text{S}=\sum_{k=0}^{N}
  D_{\frac{N}{2}-k,\frac{N}{2}}^{(\frac{N}{2})}(\varphi,\vartheta,\chi)\left|{\textstyle N-k},k\right\rangle,\label{conditional3}
\end{equation}
where we have used Eq.~(\ref{quasito}) and Fock-state notation on the r.h.s.
This is precisely a state of the form given in Eq.~(\ref{quasiBell}) with coefficients
determined by the Wigner rotation matrix elements. Since these elements depend solely on the coupling
constants, the generated entangled state is sensitive to their
magnitudes and phases. 
The state in Eq.~(\ref{conditional3}) corresponds to the quasi-mode Fock state
$\dleft|N,0\drangle$, implying that each conditional step
generates one photon in quasi-mode one.
The generation probabilities of the states $|\Psi_N\rangle$ and $|\Psi_N^\pm\rangle$ after $N$
conditional steps are given by
\begin{eqnarray}\label{conditional2}
  \left|\left\langle\Psi_N\mid\chi_N\right\rangle\right|^2&=&\left|\sum_{m=-\frac{N}{2}}^{N/2}\tilde{c}^\ast_{\frac{N}{2}m}D_{m\frac{N}{2}}^{(\frac{N}{2})}\right|^2,\\
 \left|\left\langle\Psi_N^\pm\mid\chi_N\right\rangle\right|^2&=&\frac{1}{2}\left|D_{\frac{N}{2}\frac{N}{2}}^{(\frac{N}{2})}\pm D_{-\frac{N}{2}\frac{N}{2}}^{(\frac{N}{2})}\right|^2. 
\end{eqnarray}
We shall show that the probability to detect the atoms $N$ times consecutively
in the ground state is a rapidly decaying function of $N$.
But, as discussed below, it is not essential to rely on this
assumption. Actually, it is sufficient to detect them in a sequence of $n(\ge N)$ steps $N$ times in the
ground state.
\section{Non-Conditional Generation}
\par In the following we consider a {\em non-conditional} scheme.
We start with an empty cavity and send a sequence of excited
atoms through it without detecting their final states. 
The reduced density operator of the field after the passage of the first atom (interaction time
$\tau_1$) is given by
\begin{eqnarray}\label{uncond1}
  \rho_{\text{F}}^{(1)}(\tau_1)
  &=&\cos^2{(\tau_1)}\dleft|0,0\drangle_\text{S}\!\dlangle0,0\dright| +
  \sin^2{(\tau_1)}\dleft|{\textstyle\frac{1}{2},\frac{1}{2}}\drangle_\text{S}\!\dlangle{\textstyle\frac{1}{2},\frac{1}{2}}\dright|\nonumber
\end{eqnarray}
and serves as the ``initial'' field configuration for the second excited atom.
Proceeding this way, the reduced density operator of the field after $n$
steps turns out to be of the form
\begin{equation}
   \rho_{\text{F}}^{(n)}(\{\tau_n\})=\mathop{\sum\nolimits'}\limits_{j=0}^{n/2}
  p_j^{(n)}(\{\tau_n\}){{\dleft|j,j\drangle}_\text{S}\!\dlangle j,j
  \dright|},\label{rhonc}
\end{equation}
where the coefficients $p_j^{(n)}$ are given recursively by
\begin{eqnarray}
  p_0^{(n)}&=&\cos^2{\left(\tau_{n}\right)}p_0^{(n-1)},\nonumber\\
  p_j^{(n)}&=&\cos^2{\left(\tau_{n}\sqrt{2j+1}\right)}p_j^{(n-1)}+\sin^2{\left(\tau_{n}\sqrt{2j}\right)}p_{j-\frac{1}{2}}^{(n-1)},\nonumber\\
  p_{n/2}^{(n)}&=&\sin^2{\left(\tau_{n}\sqrt{n}\right)}p_{(n-1)/2}^{(n-1)},\label{recursion}
\end{eqnarray}
for $1/2\le j\le (n-1)/2$ and $p_0^{(0)}=1$. The argument $\{\tau_n\}$ stands
for all interaction times $(\tau_1,\ldots,\tau_n)$ of the $n$ steps.
Equation~(\ref{rhonc}), which is obviously true for $n=1$ and $n=2$ (see Eq.~(\ref{uncond2})), can be
proven by induction.
\par 
The coefficients $p_j^{(n)}$ in Eq.~(\ref{rhonc}) are the probabilities to
find the field after $n$ non-conditional steps in the quasi-mode state $\dleft|j,j\drangle_\text{S}$. 
In particular
$p_0^{(n)}=\cos^2{(\tau_1)}\cos^2{(\tau_2\sqrt{2})}\allowbreak\ldots\allowbreak\cos^2{(\tau_n\sqrt{n})}$
and 
$p_{n/2}^{(n)}=\sin^2{(\tau_1)}\sin^2{(\tau_2\sqrt{2})}\allowbreak\ldots\allowbreak\sin^2{(\tau_n\sqrt{n})}$
correspond to the cases where in $n$ steps the initially excited atoms emerge $n$ times in the excited and ground state, respectively. The
intermediate $p_j^{(n)}$'s correspond to the cases where the $n$ atoms emerge $2j$ times in the ground state and $n-2j$ times in the excited state,
irrespective of the order of appearance.
The coefficient $p_j^{(n)}$ consists of a sum of $\binom{n}{2j}$
terms, each of which corresponds to a particular sequence of $|g\rangle$'s and
$|e\rangle$'s contributing, respectively, a sine squared and cosine squared factor. There
are altogether $2^n$ terms in Eq.~(\ref{rhonc}).  
All this is easily seen by giving $\rho_\text{F}^{(2)}$ as an example:
\begin{eqnarray}\label{uncond2}
  \lefteqn{\rho_\text{F}^{(2)}=\cos^2{\tau_1}\cos^2{\tau_2}\dleft|0,0\drangle_\text{S}\!\dlangle
  0,0\dright|}\nonumber\\
  &&+\left(\cos^2{\tau_1}\sin^2{\tau_2}+\sin^2{\tau_1}\cos^2{(\tau_2\sqrt{2})}\right)
    \dleft|\textstyle{\frac{1}{2},\frac{1}{2}}\drangle_\text{S}\!\dlangle\textstyle{\frac{1}{2},\frac{1}{2}}\dright|\nonumber\\
  &&+\sin^2{\tau_1}\sin^2{(\tau_2\sqrt{2})}\dleft|1,1\drangle_\text{S}\!\dlangle 1,1\dright|.
\end{eqnarray}
Here the four terms correspond to the final state sequences ($e$,$e$), ($e$,$g$), ($g$,$e$), and ($g$,$g$).
\par The states $|\Psi_N\rangle$ and $|\Psi_N^\pm\rangle$ are generated in a non-conditional $n$-step process with probabilities
\begin{eqnarray}
  \left\langle\rho_{\Psi_N}\right\rangle&=&p_\frac{N}{2}^{(n)}\left|\sum_{m=-\frac{N}{2}}^{N/2}\tilde{c}_{\frac{N}{2}m}^\ast
  D_{m\frac{N}{2}}^{(\frac{N}{2})}\right|^2,\\
\left\langle\rho_{\Psi_N^\pm}\right\rangle&=&\frac{1}{2}p_\frac{N}{2}^{(n)}\left|D_{\frac{N}{2}\frac{N}{2}}^{(\frac{N}{2})}\pm D_{-\frac{N}{2}\frac{N}{2}}^{(\frac{N}{2})}\right|^2, 
\end{eqnarray}
which are the {\em conditional} probabilities found before, multiplied by the
probability $p_{N/2}^{(n)}$.
Here the interaction times must be chosen such that
$p_{N/2}^{(n)}\not=0$, which amounts to control the $n$ parameters $(\tau_1,\ldots,\tau_n)$.
The state $|\Psi_N\rangle$ can be generated in a minimum number of $n=N$
steps with probability $p_{N/2}^{(N)}$ which, however is a rapidly decaying function
of $N$.  
\par In the non-conditional scheme all field states $\dleft|j,j\drangle_\text{S}=\dleft|2j,0\drangle$ for
$j=0,1/2,\ldots,n/2$ are produced, Eq.~(\ref{rhonc}). On the contrary, in the {\em conditional} scheme only
the entangled $N$-photon state $\dleft|N,0\drangle$, Eq.~(\ref{conditional3}), is generated, if in $n$ steps $N$ atoms are detected in the ground state. 
To produce $|\Psi_N\rangle$ it is, therefore, not
crucial that the atoms have been detected $N$ times consecutively in their ground state. Any
sequence of ground and excited states containing $N$ times the ground state
will do it. Finally, we note that there is a particular choice of the
interaction time of the $\ell$th atom,  given by $\tau_\ell=\pi/(2\sqrt{\ell})$ for which both, the
conditional and non-conditional scheme give (with probability one) the same entangled
state $\dleft|N,0\drangle$ in Eq.~(\ref{conditional3}). 
\section{Conclusion}
\par To conclude, we have solved the two-mode JC model algebraically by reducing it
to an effective one quasi-mode JC model. The mode and quasi-mode
picture are unitarily related by an SU(2) transformation. 
The solution found is used to discuss three different schemes for the
generation of entangled states of the two field modes. To generate entangled $N$-photon
states in a single step the initial state must contain at least $N-1$
photons and an excited atom. Starting from the vacuum we need at least $n\ge
N$ steps to produce pure (mixed) field states in the conditional (non-conditional) scheme presented. 
\begin{acknowledgments}
C.W. acknowledges helpful discussions with H.D.~Dahmen and
thanks R.J.~Glauber and H.~Walther for their interest and encouragement expressed at ICAP 2002.
\end{acknowledgments}

\end{document}